# Multivariate Functional Regression Models for Epistasis Analysis

Futao Zhang[1], Dan Xie[2], Meimei Liang[3], Momiao Xiong[4,*]

[1] Department of Computer Science, College of Internet of Things, Hohai University, Changzhou, 213022, China

[2] College of Information Engineering, Hubei University of Chinese Medicine, Hubei, 430065, China

[3] Institute of Bioinformatics, Zhejiang University, Hangzhou, Zhejiang, 310058, China.

[4] Human Genetics Center, Division of Biostatistics, The University of Texas School of Public Health, Houston, TX 77030, USA.

**Running Title**: Interaction Analysis

**Key words**: Gene-gene interaction, multivariate functional regression, next-generation sequencing, association studies, pleiotropic interaction effect, QTL

[*]Address for correspondence and reprints: Dr. Momiao Xiong, Human Genetics Center, The University of Texas Health Science Center at Houston, P.O. Box 20186, Houston, Texas 77225, (Phone): 713-500-9894, (Fax): 713-500-0900, E-mail: Momiao.Xiong@uth.tmc.edu

## Abstract

To date, most genetic analyses of phenotypes have focused on analyzing single traits or, analyzing each phenotype independently. However, joint epistasis analysis of multiple complementary traits will increase statistical power, and hold the key to understanding the complicated genetic structure of the complex diseases. Despite their importance in uncovering the genetic structure of complex traits, the statistical methods for identifying epistasis in multiple phenotypes remains "fundamentally unexplored ". To fill this gap, we formulate a test for interaction between two gens in multiple quantitative trait analysis as a multiple functional regression (MFRG) in which the genotype functions (genetic variant profiles) are defined as a function of the genomic position of the genetic variants. We use large scale simulations to calculate its type I error rates for testing interaction between two genes with multiple phenotypes and to compare its power with multivariate pair-wise interaction analysis and single trait interaction analysis by a single variate functional regression model. To further evaluate its performance, the MFRG for epistasis analysis is applied to five phenotypes and exome sequence data from the NHLBI's Exome Sequencing Project (ESP) to detect pleiotropic epistasis. A total of 136 pairs of genes that formed a genetic interaction network showed significant evidence of epistasis influencing five traits. The results demonstrate that the joint interaction analysis of multiple phenotypes has much higher power to detect interaction than the interaction analysis of single trait and may open a new direction to fully uncovering the genetic structure of multiple phenotypes.

## Author Summary

Most genetic analyses of complex traits have focused on a single trait association analysis, analyzing each phenotype independently, and additive model in which genetic variation is assumed to contribute independently, additively and cumulatively to the trait. However, multiple phenotypes are correlated. Complex traits are influenced by many genetic and environmental factors and their interactions. Joint gene-gene (GxG) interaction analysis of multiple complementary traits will increase statistical power to identify GxG interactions, and hold the key to understanding the complicated genetic structure of the complex diseases. Despite their importance in uncovering the genetic structure of complex traits, the statistical methods for identifying GxG interactions in multiple phenotypes remains less developed owing to its potential complexity. Therefore, we propose to develop a multiple functional regression (MFRG) model in which the genotype functions (genetic variant profiles) are defined as a function of the genomic position of the genetic variants for simultaneous GxG interaction analysis of multiple correlated phenotypes. We use large scale simulations to calculate its type I error rates for testing interaction between two genes with multiple phenotypes and to compare its power with multivariate pair-wise interaction analysis and single trait interaction analysis by a single variate functional regression model. To further evaluate its performance, the MFRG for epistasis analysis is applied to five phenotypes and exome sequence data from the NHLBI's Exome Sequencing Project (ESP) to detect pleiotropic epistasis. A total of 136 pairs of genes that formed a genetic interaction network showed significant evidence of epistasis influencing five traits.

## Introduction

In the past several years we have witnessed remarkable progresses in the development of methodologies for identification of epistasis which detect deviation from summation of genetic additive effects for a quantitative trait [1]. The methods for epistasis analysis can be divided into two categories: SNP-based and group-based interaction analysis. SNP-based methods test for all pairwise interactions between SNPs, while group-based methods detect interactions between groups of SNPs. Regression-based methods [2-8], haplotype-based methods [9-15], machine learning-based methods [16-20] are widely used for epistasis analysis.

To date, most genetic analyses of phenotypes have focused on analyzing single traits or, analyzing each phenotype independently [21]. However, multiple phenotypes are highly correlated. It has been reported that more than 4.6% of the SNPs and 16.9% of the genes in previous genome-wide association studies (GWAS) were significantly associated with more than one trait [22]. These results demonstrate that genetic pleiotropic effects likely play a crucial role in the molecular basis of correlated phenotypes [23-26]. Joint epistasis analysis of multiple complementary traits will increase statistical power to identify epistasis, and hold the key to understanding the complicated genetic structure of the complex diseases [27, 28]. Despite their importance in uncovering the genetic structure of complex traits, the statistical methods for identifying epistasis in multiple phenotypes remains "fundamentally unexplored owing to its potential complexity" [1]. The interaction analysis for multiple phenotypes have been limited to common variants in carefully controlled experimental crosses [29, 30]. Simultaneously analyzing interactions for multiple phenotypes in humans poses enormous challenges for methodologies and computations.

Purpose of this paper is to develop a general analytic framework and novel statistical methods for simultaneous epistasis analysis of multiple correlated phenotypes. To unify approach to epistasis analysis for both common and rare variants, we take a genome region (or gene) as a basic unit of interaction analysis and use all the information that can be accessed to collectively test interaction between all possible pairs of SNPs within two genome regions (or genes). We use the functional data analysis to reduce the dimension of NGS data. Specifically, we use genetic variant profiles which will recognize information contained in the physical location of the SNP as a major data form. The densely typed genetic variants in a genomic region for each individual are so close that these genetic variant profiles can be treated as observed data taken from curves [8, 31]. Since standard multivariate statistical analyses often fail with functional data [32] we formulate a test for interaction between two genomic regions in multiple quantitative trait analysis as a multiple functional regression (MFRG) model [33] with scalar response. In the MFRG model the genotype functions (genetic variant profiles) are defined as a function of the genomic position of the genetic variants rather than a set of discrete genotype values and the quantitative trait is predicted by genotype functions with their interaction terms. By functional principal component analysis, the genotype functions are expanded as a few functional principal components (FPC) and the MFRG model is transformed to the classical multivariate regression model (MRG) in which FPC scores are taken as variates. The develop statistics in the paper can be applied to pair-wise interaction tests and gene-based interaction tests for multiple phenotypes. By investigating SNP-SNP interactions or gene-gene interactions that are shared across multiple traits, we can study pleiotropic epistasis.

To evaluate its performance for multitrait epistasis analysis, we use large scale simulations to calculate the type I error rates of the MFRG for testing interaction between two genomic regions with multiple phenotypes and to compare its power with multivariate pair-wise interaction analysis and single trait interaction analysis by functional regression (FRG) model. To further evaluate its performance, the MFRG for epistasis analysis is applied to five traits: high density lipoprotein (HDL), low density lipoprotein (LDL), total cholesterol, systolic blood pressure (SBP) and diastolic blood pressure (DBP), and exome sequence data from the NHLBI's Exome Sequencing Project (ESP) to detect pleiotropic epistasis.

## Method

Assume that $n$ individuals are sampled. Let $y_{ik}, k = 1,2,...,K$ be $K$ trait values of the $i$ -th individual. Consider two genomic regions $[a_1, b_1]$ and $[a_2, b_2]$. Let $x_i(t)$ and $x_i(s)$ be genotypic functions of the $i$-th individual defined in the regions $[a_1, b_1]$ and $[a_2, b_2]$, respectively. Let $y_i$ be the phenotypic value of a quantitative trait measured on the $i$ -th individual. Let $t$ and $s$ be a genomic position in the first and second genomic regions, respectively. Define a genotype profile $x_i(t)$ of the i-th individual as

$$X_i(t) = \begin{cases} 0, & \text{mm} \\ 1, & \text{Mm} \\ 2, & \text{MM}, \end{cases}$$

where M and m are two alleles of the marker at the genomic position *t*. Recall that a regression model for interaction analysis with the $k$ -th trait is defined as

$$y_{ik} = \mu_k + \sum_{j=1}^{J_1} x_{ij}\alpha_{kj} + \sum_{l=1}^{J_2} z_{il}\beta_{kl} + \sum_{j=1}^{J_1}\sum_{l=1}^{J_2} x_{ij}z_{il}\gamma_{kjl} + \varepsilon_{ik}, \qquad (1)$$

where $\mu_k$ is an overall mean of the $k$ -th trait, $\alpha_{kj}$ is the main genetic additive effect of the $j$ -th SNP in the first genomic region for the $k$ -th trait, $\beta_{kl}$ is the main genetic additive effect of the $l$ -th SNP in the second genomic region for the $k$ -th trait, $\gamma_{kjl}$ is an additive $\times$ additive interaction effect between the $j$ -th SNP in the first genomic region and the $l$ -th SNP in the second genomic region for the $k$ -th trait, $x_{ij}$ and $z_{il}$ are indicator variable for the genotypes at the $j$ -th SNP and the $l$ -th SNP, respectively, $\varepsilon_{ik}, k = 1,..,K$ are independent and identically distributed normal variables with mean of zero and covariance matrix $\Sigma$ .

Similar to the multiple regression models for interaction analysis with multiple quantitative traits, the functional regression model for a quantitative trait can be defined as

$$y_{ik} = \alpha_{0k} + \int_T \alpha_k(t)x_i(t)dt + \int_S \beta_k(s)x_i(s)ds + \int_T \int_S \gamma_k(t,s)x_i(t)x_i(s)dtds + \varepsilon_{ik}, \quad (2)$$

where $\alpha_{0k}$ is an overall mean, $\alpha_k(t)$ and $\beta_k(s)$ are genetic additive effects of two putative QTLs located at the genomic positions $t$ and $s$ , respectively, $\gamma_k(t,s)$ is the interaction effect between two putative QTLs located at the genomic positions $t$ and $s$ for the $k$ -th trait, $k = 1,...,K$ , $x_i(t)$ and $x_i(s)$ are genotype profiles, $\varepsilon_{ik}$ are independent and identically distributed normal variables with mean of zero and covariance matrix $\Sigma$ .

**Estimation of Interaction Effects**

We assume that both phenotypes and genotype profiles are centered. The genotype profiles $x_i(t)$ and $x_i(s)$ are expanded in terms of the orthonormal basis function as:

$$x_i(t) = \sum_{j=1}^{\infty} \xi_{ij}\phi_j(t) \text{ and}$$

$$x_i(s) = \sum_{l=1}^{\infty} \eta_{il}\psi_l(s) , \quad (3)$$

where $\phi_j(t)$ and $\psi_l(s)$ are sequences of the orthonormal basis functions. The expansion coefficients $\xi_{ij}$ and $\eta_{il}$ are estimated by

$$\xi_{ij} = \int_T x_i(t)\phi_j(t)dt \text{ and}$$

$$\eta_{il} = \int_S x_i(s)\psi_l(s)ds\,. \tag{4}$$

In practice, numerical methods for the integral will be used to calculate the expansion coefficients.

Substituting equation (3) into equation (2), we obtain

$$\begin{aligned} y_{ik} &= \int_T \alpha_k(t)\sum_{j=1}^{\infty}\xi_{ij}\phi_j(t)dt + \int_S \beta_k(s)\sum_{l=1}^{\infty}\eta_{il}\psi_l(s)ds \int_T\int_S \gamma(t,s)(\sum_{j=1}^{\infty}\xi_{ij}\phi_j(t))(\sum_{l=1}^{\infty}\eta_{il}\psi_l(s))dtds + \varepsilon_i \\ &= \sum_{j=1}^{\infty}\xi_{ij}\int_T \alpha_k(t)\phi_j(t)dt + \sum_{l=1}^{\infty}\eta_{il}\int_S \beta_k(s)\psi_l(s)ds + \sum_{j=1}^{\infty}\sum_{l=1}^{\infty}\xi_{ij}\eta_{il}\int_T\int_S \gamma_k(t,s)\phi_j(t)\psi_l(s)dtds + \varepsilon_{ik} \quad (5) \\ &= \sum_{j=1}^{\infty}\xi_{ij}\alpha_{kj} + \sum_{l=1}^{\infty}\eta_{il}\beta_{kl} + \sum_{j=1}^{\infty}\sum_{l=1}^{\infty}\xi_{ij}\eta_{il}\gamma_{kjl} + \varepsilon_{ik}, i = 1,\ldots,n, k = 1,\ldots,K, \end{aligned}$$

where $\alpha_{kj} = \int_T \alpha_k(t)\phi_j(t)dt$, $\beta_{kl} = \int_S \beta_k(s)\psi_l(s)ds$, and $\gamma_{kjl} = \int_T\int_S \gamma_k(t,s)\phi_j(t)\psi_l(s)dtds$. The parameters $\alpha_{kj}, \beta_{kl}$ and $\gamma_{kjl}$ are referred to as genetic additive and additive $\times$ additive effect scores for the $k$ -th trait. These scores can also be viewed as the expansion coefficients of the genetic effect functions with respect to orthonormal basis functions:

$$\alpha_k(t) = \sum_j \alpha_{kj}\phi_j(t), \beta_k(s) = \sum_l \beta_{kl}\psi_l(s) \text{ and } \gamma_k(s,t) = \sum_j\sum_l \gamma_{kjl}\phi_j(s)\psi_l(t)\,. \tag{6}$$

Let

$$Y=[Y_1,\ldots,Y_K]=\begin{bmatrix} Y_{11} & \cdots & Y_{1K} \\ \vdots & \ddots & \vdots \\ Y_{n1} & \cdots & Y_{nK} \end{bmatrix},\ \xi=\begin{bmatrix} \xi_{11} & \cdots & \xi_{1J} \\ \vdots & \ddots & \vdots \\ \xi_{n1} & \cdots & \xi_{nJ} \end{bmatrix}, \eta=\begin{bmatrix} \eta_{11} & \cdots & \eta_{1L} \\ \vdots & \ddots & \vdots \\ \eta_{n1} & \cdots & \eta_{nL} \end{bmatrix},\quad \xi_i=\begin{bmatrix} \xi_{i1} \\ \vdots \\ \xi_{iJ} \end{bmatrix},$$

$$\eta_i=\begin{bmatrix} \eta_{i1} \\ \vdots \\ \eta_{iL} \end{bmatrix}, \Gamma=\begin{bmatrix} \xi_1^T \otimes \eta_1^T \\ \vdots \\ \xi_n^T \otimes \eta_n^T \end{bmatrix}=\begin{bmatrix} \xi_{11}\eta_{11} & \cdots & \xi_{11}\eta_{1L} & \cdots & \xi_{1J}\eta_{11} & \cdots & \xi_{1J}\eta_{1L} \\ \cdots & \cdots & \cdots & \cdots & \cdots & \cdots & \cdots \\ \xi_{n1}\eta_{n1} & \cdots & \xi_{n1}\eta_{nL} & \cdots & \xi_{nJ}\eta_{n1} & \cdots & \xi_{nJ}\eta_{nL} \end{bmatrix}, \gamma_k=\begin{bmatrix} \gamma_{k11} \\ \vdots \\ \gamma_{k1L} \\ \gamma_{k21} \\ \vdots \\ \gamma_{k2L} \\ \vdots \\ \gamma_{kJ1} \\ \vdots \\ \gamma_{kJL} \end{bmatrix},$$

$$\gamma=\begin{bmatrix}\gamma_1 & \cdots & \gamma_K\end{bmatrix}, \alpha_k=\begin{bmatrix} \alpha_{k1} \\ \vdots \\ \alpha_{kJ} \end{bmatrix}, \alpha=[\alpha_1,\ldots,\alpha_K], \beta_k=\begin{bmatrix} \beta_{k1} \\ \vdots \\ \beta_{kL} \end{bmatrix}, \beta=[\beta_1,\ldots,\beta_K], \varepsilon=\begin{bmatrix} \varepsilon_{11} & \cdots & \varepsilon_{1K} \\ \cdots & \cdots & \cdots \\ \varepsilon_{n1} & \cdots & \varepsilon_{nK} \end{bmatrix}.$$

Then, equation (5) can be approximated by

$$\begin{aligned} Y &= \xi\alpha+\eta\beta+\Gamma\gamma+\varepsilon \\ &= WB+\varepsilon, \end{aligned} \tag{7}$$

where $W=\begin{bmatrix}\xi & \eta & \Gamma\end{bmatrix}$ and $B=\begin{bmatrix}\alpha \\ \beta \\ \gamma\end{bmatrix}$.

The standard least square estimators of B and the variance covariance matrix $\Sigma$ are, respectively, given by

$$\hat{B}=(W^TW)^{-1}W^T(Y-\bar{Y}), \tag{8}$$

$$\hat{\Sigma}=\frac{1}{n}(Y-WB)^T(Y-WB). \tag{9}$$

Denote the last $JL$ row of the matrix $(W^T W)^{-1} W^T$ by $A$. Then, the estimator of the parameter $\gamma$ is given by

$$\hat{\gamma} = A(Y - \overline{Y}). \quad (10)$$

The vector of the matrix $\gamma$ can be written as

$$vec(\hat{\gamma}) = (A \otimes I) vec(Y - \overline{Y}). \quad (11).$$

By the assumption of the variance matrix of $Y$, we obtain the variance matrix of $vec(Y)$:

$$\text{var}(vec(Y)) = \Sigma \otimes I. \quad (12)$$

Thus, it follows from equations (11) and (12) that

$$\begin{aligned} \Lambda = \text{var}(vec(\tilde{\gamma})) &= (A \otimes I)(\Sigma \otimes I)(A^T \otimes I) \\ &= (A \Sigma A^T) \otimes I. \end{aligned} \quad (13)$$

**Test Statistics**

An essential problem in genetic interaction studies of the quantitative traits is to test the **interaction between two genomic regions (or genes). Formally, we investigate the problem of** testing the following hypothesis:

$$\gamma_k(t,s) = 0, \forall t \in [a_1, b_1], s \in [a_2, b_2], k = 1, ..., K, ,$$

which is equivalent to testing the hypothesis:

$$H_0 : \gamma = 0.$$

Define the test statistic for testing the interaction between two genomic regions $[a_1, b_1]$ and $[a_2, b_2]$ with $K$ quantitative traits as

$$T_I = (vec(\hat{\gamma})^T \Lambda^{-1} vec(\hat{\gamma}) . \quad (14)$$

Then, under the null hypothesis $H_0 : \gamma = 0$ , $T_I$ is asymptotically distributed as a central $\chi^2_{(KJL)}$ distribution if $JL$ components are taken in the expansion equation (3).

We can also develop likelihood ratio-based statistics for testing interaction.

Setting $W = [W_1 \quad W_2]$, we can write the model as

$$E[Y] = W_1 \begin{bmatrix} \alpha \\ \beta \end{bmatrix} + W_2 \gamma .$$

Under $H_0 : \gamma = 0$, we have the model:

$$Y = W_1 \begin{bmatrix} \alpha \\ \beta \end{bmatrix} + \varepsilon.$$

The estimators will be

$$\begin{bmatrix} \hat{\alpha} \\ \hat{\beta} \end{bmatrix} = (W_1^T W_1)^{-1} W_1^T Y \text{ and } \hat{\Sigma}_1 = \frac{1}{n} (Y - W_1 \begin{bmatrix} \hat{\alpha} \\ \hat{\beta} \end{bmatrix})^T (Y - W_1 \begin{bmatrix} \hat{\alpha} \\ \hat{\beta} \end{bmatrix}) .$$

The likelihood for the full model and reduced model are, respectively, given by

$$L(\hat{\alpha}, \hat{\beta}, \hat{\gamma}, \hat{\Sigma}) = \frac{e^{-nK/2}}{(2\pi)^{nK/2} | \hat{\Sigma} |^{n/2}} \text{ and}$$

$$L(\hat{\alpha}, \hat{\beta}, \hat{\Sigma}_1) = \frac{e^{-nK/2}}{(2\pi)^{nK/2} | \hat{\Sigma}_1 |^{n/2}} .$$

The likelihood-ratio-based statistic for testing interaction between two genomic regions with multivariate traits is defined as

$$T_{I\Lambda} = -n \log \left( \frac{| \hat{\Sigma} |}{| \hat{\Sigma}_1 |} \right) . \quad (15)$$

Under the null hypothesis $H_0 : \gamma = 0$ , $T_{I\Lambda}$ is asymptotically distributed as a central $\chi^2_{(KJL)}$ distribution if $JL$ components are taken in the expansion equation (3).

**Results**

**Null Distribution of Test Statistics**

To examine the null distribution of test statistics, we performed a series of simulation studies to compare their empirical levels with the nominal ones. We calculated the type I error rates for rare alleles, and common alleles. We first assumed the model with no marginal effects for all traits:

$$Y_i = \mu + \varepsilon_i \text{, } i = 1,...,n,$$

where $Y_i = [y_{i1},..., y_{ik}], \mu = [\mu_{1,} ..., \mu_k]$, and $\varepsilon_i$ is distributed as

$$[\varepsilon_1 \quad ... \quad \varepsilon_k] \sim N\left([0 \quad ... \quad 0], \begin{pmatrix} 1 & \cdots & 0.5 \\ \vdots & \ddots & \vdots \\ 0.5 & \cdots & 1 \end{pmatrix}\right)$$

Then, we considered the model with marginal genetic effect (additive model) at one gene:

$$y_{ik} = \mu_k + \sum_{j=1}^{J} x_{ij}\alpha_{kj} + \varepsilon_{ik},$$

where

$$x_{ij} = \begin{cases} 2(1-P_j) & A_j A_j \\ 1-2P_j & A_j a_j \\ -2P_j & a_j a_j \end{cases}, \quad \alpha_k = (r_k - 1) f_0,$$

$P_j$ is a frequency of the allele $A_j$ , $r_k$ is a risk parameter of the $k$ -th trait which was randomly selected from 1.1 to 1.6, $f_0$ is a baseline penetrance and set to 1 and $\varepsilon$ are defined as before.

Finally, we consider the model with marginal genetic effects (additive model) at both genes:

$$y_{ik} = \mu_k + \sum_{j=1}^{J} x_{ij}\alpha_{kj} + \sum_{l=1}^{L} z_{il}\beta_{kl} + \varepsilon_{ik},$$

where

$$x_{ij}=\begin{cases}2(1-P_j) & A_jA_j\\ 1-2P_j & A_ja_j\\ -2P_j & a_ja_j\end{cases},\quad \mathrm{z}_{il}=\begin{cases}2(1-q_l) & B_lB_l\\ 1-2q_l & B_lb_l\\ -2q_l & b_lb_l\end{cases},\ \alpha_{kj}=\alpha_k=(r_{pk}-1)f_0,\beta_{kl}=\beta_k=(r_{qk}-1)f_0,$$

$P_j$ and $q_l$ are frequencies of the alleles $A_j$ and $B_l$, respectively, $r_{pk}$ and $r_{qk}$ are risk parameters of the $k$-th trait for the SNPs in the first and second genes, respectively, and randomly selected from 1.1 to 1.6, $f_0$ is a baseline penetrance and set to 1 and $\varepsilon$ are defined as before.

We generated 1,000,000 chromosomes by resampling from 2,018 individuals with variants in five genes (*TNFRSF14*,*GBP3*,*KANK4*, *IQGAP3*, *GALNT2*) selected from the NHLBI's Exome Sequencing Project (ESP). We randomly selected 20% of SNPs as causal variants. The number of sampled individuals from populations of 1,000,000 chromosomes ranged from 1,000 to 5,000. We presented average type 1 error rates over 10 pairs of genes selected from the above five genes. A total of 5,000 simulations were repeated.

Table 1 and supplemental Tables S1 and S2 summarized the average type I error rates of the test statistics for testing the interaction between two genes with no marginal effect consisting only rare variants with 5 traits， 2 and 10 traits, respectively, over 10 pairs of genes at the nominal levels α=0.05, α=0.01 and α=0.001. Table 2 and supplemental Tables S3 and S4 summarized the average type I error rates of the test statistics for testing the interaction between two genes with marginal effect at one gene consisting only rare variants with 5 traits， 2 and 10 traits, respectively, over 10 pairs of genes at the nominal levels α=0.05, α=0.01 and α=0.001. Table 3 and supplemental Tables S5 and S6 summarized the average type I error rates of the test statistics for testing the interaction between two genes with marginal effect at both genes consisting only rare variants with 5 traits, 2 and 10 traits, respectively, over 10 pairs of genes at the nominal levels α=0.05, α=0.01 and α=0.001. For common variants, we summarized the average type I error rates of the test statistics for testing the interaction between two genes with marginal effect at both genes consisting only common variants with 5 traits, 2 and 10 traits, respectively, over 10 pairs of genes at the nominal levels α=0.05, α=0.01 and α=0.001, in Table 4 and supplemental Tables S7 and S8, respectively. The statistics for testing interaction between two genomic regions with only common variants have the similar type 1 error rates in other two scenarios: with marginal genetic effect at one gene or without marginal genetic effects at two genes (data not shown). These results clearly showed that the type I error rates of the MFRG-based test statistics for testing interaction between two genes with multiple traits with or without marginal effects were not appreciably different from the nominal $\alpha$ levels.

Table 1. Average type 1 error rates of the statistic for testing interaction between two genes with no marginal effect consisting only rare variants with 5 traits over 10 pairs of genes.

| Sample Size | 0.05 | 0.01 | 0.001 |
|---|---|---|---|
| 1000 | 0.0635 | 0.0137 | 0.0015 |
| 2000 | 0.0535 | 0.0111 | 0.0010 |
| 3000 | 0.0508 | 0.0105 | 0.0011 |
| 4000 | 0.0495 | 0.0094 | 0.0010 |
| 5000 | 0.0476 | 0.0095 | 0.0012 |

Table 2. Average type 1 error rates of the statistic for testing interaction between two genes with marginal effect at one gene consisting only rare variants with 5 traits over 10 pairs of genes.

| Sample Size | 0.05 | 0.01 | 0.001 |
|---|---|---|---|
| 1000 | 0.0641 | 0.0137 | 0.0017 |
| 2000 | 0.0556 | 0.0120 | 0.0013 |
| 3000 | 0.0499 | 0.0104 | 0.0011 |
| 4000 | 0.0490 | 0.0095 | 0.0010 |
| 5000 | 0.0486 | 0.0096 | 0.0009 |

Table 3. Average type 1 error rates of the statistic for testing interaction between two genes with marginal effects at two genes consisting only rare variants with 5 traits over 10 pairs of genes.

| Sample Size | 0.05 | 0.01 | 0.001 |
|---|---|---|---|
| 1000 | 0.0651 | 0.0139 | 0.0017 |
| 2000 | 0.0526 | 0.0111 | 0.0013 |
| 3000 | 0.0521 | 0.0105 | 0.0010 |
| 4000 | 0.0486 | 0.0093 | 0.0009 |
| 5000 | 0.0498 | 0.0096 | 0.0011 |

Table 4. Average type 1 error rates of the statistic for testing interaction between two genes with marginal effects at two genes consisting only common variants with 5 traits over 10 pairs of genes.

| Sample size | 0.05 | 0.01 | 0.001 |
|---|---|---|---|
| 1000 | 0.0536 | 0.0119 | 0.0016 |
| 2000 | 0.0502 | 0.0100 | 0.0010 |
| 3000 | 0.0478 | 0.0093 | 0.0009 |
| 4000 | 0.0477 | 0.0091 | 0.0010 |
| 5000 | 0.0484 | 0.0101 | 0.0009 |

**Power Evaluation**

To evaluate the performance of the MFRG models for interaction analysis of multiple traits, we used simulated data to estimate their power to detect interaction between two genes for two, four, five, six and ten quantitative traits. A true multiple quantitative genetic model is given as follows. Consider $H$ pairs of quantitative trait loci (QTL) from two genes (genomic regions). Let $Q_{h_1}$ and $q_{h_1}$ be two alleles at the first QTL, and $Q_{h_2}$ and $q_{h_2}$ be two alleles at the second QTL, for the $H$ pair of QTLs. Let $u_{ijkl}$ be the genotypes of the $u$ th individual with

$ij = Q_{h_1}Q_{h_1}, Q_{h_1}q_{h_1}, q_{h_1}q_{h_1}$ and $kl = Q_{h_2}Q_{h_2}, Q_{h_2}q_{h_2}, q_{h_2}q_{h_2}$, and $g_{mu_{ijkl}}$ be its genotypic value for the $m$-th trait. The following multiple regression is used as a genetic model for the $m$-th quantitative trait:

$$y_{mu} = \sum_{h=1}^{H} g_{mu_{ijkl}}^{h} + \varepsilon_{mu}, \; u = 1,2,...,n, m = 1,...,M ,$$

where $g_{mu_{ijkl}}^{h}$ is a genotypic value of the $h$-th pair of QTLs for the $m$-th quantitative trait and $\varepsilon_{mu}$ are distributed as $\begin{bmatrix}\varepsilon_1 & ... & \varepsilon_{\mathrm{m}}\end{bmatrix} \sim N\left(\begin{bmatrix}0 & ... & 0\end{bmatrix}, \begin{pmatrix} 1 & \cdots & 0.5 \\ \vdots & \ddots & \vdots \\ 0.5 & \cdots & 1 \end{pmatrix}\right)$.

Four models of interactions are considered: (1) Dominant OR Dominant, (2) Dominant AND Dominant,(3) Recessive OR Recessive and (4) Threshold model (Table S9 ). Recessive AND Recessive model is excluded due to infrequency of that condition with rare variants. The parameter $r$ varies from 0 to 1.

We generate 1,000,000 individuals by resampling from 2,018 individuals of European origin with variants in two genes *IQGAP3*and *ACTN2* selected from ESP dataset. We randomly selected 20% of the variants as causal variants. A total of 2,000 individuals for the four interaction models were sampled from the populations. A total of 1,000 simulations were repeated for the power calculation.

The power of the proposed MFRG model is compared with the single trait functional regression (SFRG) model, the multitrait pair-wise interaction test and the regression on principal components (PCs). For SNPs genotypes in each genomic region principal component analysis (PCA) were performed. The number of principal components for each individual which can explain 80% of the total genetic variation in the genomic region will be selected as the variables. Specifically, the principal component score of the i-th individual in the first and second genomic regions are denoted by $x_{i1},...,x_{ik_1}$ and $z_{i1},...z_{ik_2}$, respectively. The regression model for detection of interaction for the m -th trait is then given by

$$y_{mi} = \mu_m + \sum_{j=1}^{k_1} x_{ij}\alpha_{mj} + \sum_{l=1}^{k_2} z_{il}\beta_{ml} + \sum_{j=1}^{k_1}\sum_{l=1}^{k_2} x_{ij}z_{il}\gamma_{mjl} + \varepsilon_{mi}.$$

The power of the MFRG is compared with the traditional point-wise interaction test which takes the following model:

$$y_{mi} = \mu_m + x_{i1}\alpha_{m1} + x_{i2}\alpha_{m2} + x_{i1}x_{i2}\gamma_m + \varepsilon_{mi}, i = 1,...,n, m = 1,...,M.$$

For a pair of genes, we assume that the first gene has $k_1$ SNPs, and the second gene has $k_2$ SNPs, then, the total number of all possible pairs is $k = k_1 \times k_2$. For each pair of SNPs, we calculate a statistic for testing pair-wise interaction $T_{\text{mjpair}}$. Finally, the maximum of $T_{\text{mjpair}}$:

$T_{\max} = \max(T_{1,1\text{pair}}, T_{1,2\text{pair}},..., T_{1,\text{kpair}},..., T_{M,1\text{pair}},..., T_{M,\text{kpair}})$ is computed.

Figures 1,2 and Figures S1,S2 plotted the power curves of two-trait FRG, single trait FRG, two-trait regression on PCs and two-trait pair-wise interaction tests for a quantitative trait under Dominant OR Dominant, Dominant AND Dominant, Threshold, and Recessive OR Recessive models, respectively. Two genes only include rare variants. These power curves are a function of the risk parameter at the significance level $\alpha = 0.05$. Permutations in the point-wise interaction tests were used to adjust for multiple testing. In all cases, the two-trait FRG had the highest power to detect epistasis. We observed two remarkable features. First, two-trait test had higher power than the one-trait test. Second, the two-trait FRG had the highest power among all two-trait tests.

**Fig. 1. Power curves under Dominant OR Dominant with two genes including rare variants only.**

**Fig. 2. Power curves under Dominant AND Dominant with two genes including rare variants only.**

Figures 3,4 and Figures S3,S4 plotted the power curves of two-trait FRG, single trait FRG, two-trait regression on PCs and two-trait pair-wise interaction tests for a quantitative trait under Dominant OR Dominant, Dominant AND Dominant, Threshold and Recessive OR Recessive models, respectively. Two genes only include common variants. These power curves are a function of the risk parameter at the significance level $\alpha = 0.05$. Permutations in the point-wise interaction tests were used to adjust for multiple testing. These figures showed that the power patterns of the epistasis tests for common variants were similar to that for rare variants.

**Fig. 3. Power curves under Dominant OR Dominant with two genes including common variants only.**

**Fig. 4. Power curves under Dominant AND Dominant with two genes including common variants only.**

Next we investigate the impact of the number of traits on the power. Figure 5 plotted the power curves of two-trait FRG, four-trait FRG, five-trait FRG, six-trait FRG and ten-trait FRG under Dominant OR Dominant interaction model. Figure 5 showed that if the multiple phenotypes are correlated then the power of the MFRG to detect epistasis will increase as the number of phenotypes increases.

**Fig. 5. Power curves of MFRG with different trait number under Dominant OR Dominant.**

To investigate the impact of sample size on the power, we plotted Figure 6 and Figures S5-S7 showing the power of three statistics for testing the interaction between two genomic regions (or genes) with only rare variants as a function of sample sizes under four interaction models, assuming 20 % of the risk rare variants and the risk parameter $r = 0.05$ for Dominant OR Dominant, Dominant AND Dominant, and Recessive OR Recessive, and $r = 0.5$ for Threshold models, respectively. Again, we observed that the power of the two-trait FRG was the highest. The power patterns of the four statistics to test epistasis between two regions consisting of only common variants or both common and rare variants were similar (data not shown).

**Fig. 6. Power curves of MFRG with different sample size under Dominant OR Dominant.**

**Application to Real Data Examples**

To further evaluate its performance, the MFRG for testing epistasis was applied to data from the NHLBI's ESP Project. We consider five phenotypes: HDL, LDL, total cholesterol, SBP and DBP. A total of 2,018 individuals of European origin from 15 different cohorts in the ESP Project. No evidence of cohort- and/or phenotype-specific effects, or other systematic biases was found [34]. Exomes from related individuals were excluded from further analysis. We took the rank-based inverse normal transformation of the phenotypes [35] as trait values. The total number of genes tested for interactions which included both common and rare variants was 18,498. The remaining annotated human genes which did not contain any SNPs in our dataset were excluded from the analysis. A P-value for declaring significant interaction after applying the Bonferroni correction for multiple tests was $2.92\times10^{-10}$. To examine the behavior of the MFRG, we plotted QQ plot of the two-trait FRG test (Figure 7). The QQ plots showed that the false positive rate of the MFRG for detection of interaction in some degree is controlled.

**Fig. 7. QQ plot of the two-trait FRG test for ESP dataset.**

A total of 104 pairs of genes which were derived from 85 genes showed significant evidence of epistasis with P-values < $2.9\times10^{-10}$ which were calculated using the MFRG model and simultaneously analyzing interaction of inverse normally transformed HDL and LDL (Table S10). We listed top 30 pairs of significantly interacted genes with HDL and LDL in Table 5. In Table 5 and Table S10, P-values for testing interactions between genes by regression on PCA and the minimum of P-values for testing all possible pairs of SNPs between two genes using standard regression model simultaneously analyzed for the HDL and LDL and P-values for testing epistasis by the FRG separately against single trait HDL or LDL were also listed. Several remarkable features from these results were observed. First, we observed that although pairs of genes showed no strong evidence of interactions influencing individual trait HDL or LDL, they indeed demonstrated significant interactions if interactions were simultaneously analyzed for correlated HDL and LDL. Second, the MFRG often had a much smaller P-value to detect interaction than regression on the PCA and the minimum of P-values of pair-wise tests. Third, pairs of SNPs between two genes jointly have significant interaction effects, but individually each pair of SNPs make mild contributions to the interaction effects as shown in Table 6. There were a total of 561 pairs of SNPs between genes *ST20* and *SHPK*. Table 6 listed 26 pairs of SNPs with P-values < 0.046. None of the 26 pairs of SNPs showed strong evidence of interaction. However, a number of pairs of SNPs between genes *ST20* and *SHPK* collectively demonstrated significant interaction influencing the traits HDL and LDL. Fourth, we did not observe the epistasis by individual trait analysis at the genome-wide significance level $P < 2.92 \times 10^{-10}$. However, if we release the significance level to $P < 1.5 \times 10^{-8}$, we observed five pairs of interacting genes against HDL, five pairs of interacting genes against LDL and two pleiotropic epistasis (Figure 8).Fifth, 104 pairs of interacting genes formed a network (Figure 9). The genes *C5orf64* that had interactions with 28 genes , *CSMD1* that had interactions with 25 genes, *SHPK* that had interactions with 24 genes, and *ST20* that had interactions with 18 genes were hub genes in the network. It was reported that *CSMD1* was associated with multivariate phenotype defined as low levels of low density lipoprotein cholesterol (LDL-C < or = 100 mg/dl) and high levels of triglycerides (TG > or = 180 mg/dl) [36] , associated with hypertension [37]. It was also reported that *GRIK2* and *CSMD1* interacted to influence the progression of nicotine dependence [38].

**Fig. 8. Pleiotropic epistasis of HDL and LDL.**

Five pairs of interacting gene influence variation of the HDL, five pairs of interacting gene influence variation of the LDL and two pairs of interacting genes influence the variation of both HDL and LDL.

**Fig. 9. Networks of 104 pairs of genes showing significant evidence of interactions as identified by MFRG.**

Table 5. P-values of top 30 pairs of significantly interacted genes with HDL and LDL.

| Gene 1 | Chr | Gene 2 | Chr | P-values | | | | |
|---|---|---|---|---|---|---|---|---|
| | | | | Two Traits | | | HDL | LDL |
| | | | | MFRG | Pair-wise (minimum) | PCA | FRG | FRG |
| *ST20* | 15 | *SHPK* | 17 | 4.7E-19 | 1.7E-05 | 1.4E-02 | 8.9E-09 | 2.2E-10 |
| *PDE4DIP* | 1 | *ST20* | 15 | 7.9E-19 | 1.6E-04 | 1.3E-01 | 4.9E-05 | 4.0E-09 |
| *C5orf64* | 5 | *SHPK* | 17 | 7.3E-17 | 1.7E-05 | 7.4E-03 | 1.3E-08 | 5.4E-09 |
| *C5orf64* | 5 | *ESRP2* | 16 | 9.5E-16 | 1.7E-05 | 2.0E-01 | 1.3E-05 | 1.1E-10 |
| *ADRA1B* | 5 | *CSMD1* | 8 | 2.6E-15 | 1.2E-04 | 6.6E-02 | 7.3E-09 | 1.3E-05 |
| *C5orf64* | 5 | *P4HA2* | 5 | 2.6E-15 | 5.9E-05 | 2.2E-01 | 3.0E-09 | 2.5E-07 |
| *CSMD1* | 8 | *ICE2* | 15 | 2.8E-15 | 4.9E-05 | 2.7E-02 | 1.4E-04 | 1.2E-04 |
| *MAST4* | 5 | *SHPK* | 17 | 2.7E-14 | 1.4E-05 | 2.1E-04 | 4.5E-04 | 5.3E-08 |
| *CSMD1* | 8 | *FOXO1* | 13 | 3.1E-14 | 1.6E-04 | 2.1E-02 | 1.8E-06 | 8.3E-07 |
| *PAIP2B* | 2 | *SHPK* | 17 | 4.1E-14 | 1.9E-05 | 6.3E-04 | 6.8E-06 | 2.8E-08 |
| *C5orf64* | 5 | *RNF112* | 17 | 6.0E-14 | 1.8E-04 | 2.9E-01 | 2.3E-08 | 1.1E-06 |
| *CSMD1* | 8 | *SHPK* | 17 | 1.0E-13 | 1.1E-05 | 5.1E-05 | 1.0E-03 | 2.3E-04 |
| *CSMD1* | 8 | *GSS* | 20 | 1.2E-13 | 6.2E-05 | 1.4E-01 | 4.5E-05 | 7.7E-04 |
| *LINC01588* | 14 | *SHPK* | 17 | 1.9E-13 | 1.6E-05 | 2.0E-02 | 2.0E-06 | 7.3E-08 |
| *IGHD* | 14 | *ST20* | 15 | 2.1E-13 | 5.3E-06 | 2.8E-01 | 1.1E-06 | 3.7E-06 |
| *IGHM* | 14 | *ST20* | 15 | 2.4E-13 | 5.3E-06 | 4.8E-02 | 1.5E-07 | 2.4E-05 |
| *CSMD1* | 8 | *CD300A* | 17 | 3.5E-13 | 8.5E-05 | 2.5E-01 | 8.0E-05 | 1.2E-04 |
| *CSMD1* | 8 | *C14orf37* | 14 | 3.7E-13 | 2.4E-04 | 5.2E-03 | 6.4E-05 | 1.5E-06 |
| *SHPK* | 17 | *BANF2* | 20 | 4.4E-13 | 9.4E-04 | 1.2E-03 | 2.2E-05 | 3.3E-08 |
| *PPP1R12B* | 1 | *C5orf64* | 5 | 4.8E-13 | 2.7E-04 | 2.1E-01 | 9.9E-08 | 1.3E-06 |
| *RHD* | 1 | *ST20* | 15 | 7.5E-13 | 1.0E-03 | 8.3E-02 | 7.0E-07 | 8.1E-06 |
| *PSMD1* | 2 | *C5orf64* | 5 | 7.9E-13 | 5.9E-05 | 3.2E-01 | 2.6E-05 | 8.1E-07 |
| *MEGF6* | 1 | *CSMD1* | 8 | 8.1E-13 | 2.5E-05 | 1.0E-08 | 3.0E-04 | 1.6E-06 |
| *C5orf64* | 5 | *DNAJA2* | 16 | 1.2E-12 | 4.4E-03 | 3.1E-01 | 1.2E-05 | 1.5E-08 |
| *CSMD1* | 8 | *LINC01588* | 14 | 1.7E-12 | 3.8E-04 | 1.8E-01 | 1.1E-04 | 2.9E-07 |
| *ZAP70* | 2 | *C5orf64* | 5 | 2.3E-12 | 6.1E-04 | 1.6E-01 | 1.5E-03 | 3.1E-07 |
| *CES2* | 16 | *SHPK* | 17 | 2.6E-12 | 6.3E-04 | 2.6E-01 | 9.1E-06 | 3.5E-05 |
| *CSMD1* | 8 | *PPM1A* | 14 | 2.6E-12 | 2.5E-05 | 4.5E-03 | 5.3E-06 | 1.4E-05 |
| *C5orf64* | 5 | *GSS* | 20 | 2.7E-12 | 1.1E-03 | 1.2E-01 | 1.1E-06 | 2.2E-06 |
| *CSMD1* | 8 | *CDC14B* | 9 | 3.5E-12 | 7.6E-04 | 9.8E-02 | 1.4E-05 | 1.5E-03 |

Table 6. P-values of 26 pairs of SNPs between genes *ST20* and *SHPK* for testing interaction affecting both HDL and LDL.

| Gene 1 | Gene 2 | P-value |
|---|---|---|
| *ST20* | *SHPK* | 4.66E-19 |
| *SNP1* | *SNP2* | P-Value |
| RS7257 | RS368975060 | 1.80E-02 |
| RS7257 | RS35091524 | 4.59E-02 |
| RS7257 | RS144036687 | 1.84E-05 |
| RS7257 | RS150076372 | 3.58E-02 |
| RS7257 | RS372971585 | 2.70E-03 |
| RS7257 | RS369378557 | 2.70E-03 |
| RS7257 | RS139301054 | 3.32E-02 |
| RS7257 | RS144071313 | 4.33E-02 |
| RS7257 | RS145004665 | 6.80E-03 |
| RS12440609 | RS201772890 | 1.92E-03 |
| RS12440609 | RS368975060 | 2.04E-02 |
| RS12440609 | RS141579629 | 7.00E-03 |
| RS12440609 | RS371517274 | 1.10E-02 |
| RS12440609 | RS372971585 | 2.34E-03 |
| RS12440609 | RS369378557 | 2.34E-03 |
| RS12440609 | RS201093822 | 6.61E-04 |
| RS12440609 | RS139301054 | 3.17E-02 |
| RS12440609 | RS141166207 | 6.12E-03 |
| RS189595266 | RS35091524 | 1.73E-02 |
| RS2733102 | RS368975060 | 1.84E-02 |
| RS2733102 | RS144036687 | 1.65E-05 |
| RS2733102 | RS150076372 | 3.40E-02 |
| RS2733102 | RS372971585 | 2.57E-03 |
| RS2733102 | RS369378557 | 2.57E-03 |
| RS2733102 | RS144071313 | 4.08E-02 |
| RS2733102 | RS145004665 | 7.24E-03 |

Next we analyzed five traits: HDL, LDL, SBP, DBP and TOTCHOL. Again, for each trait, inverse normal rank transformation was conducted to ensure that the normality assumption of the transformed trait variable was valid. To examine the behavior of the MFRG, we plotted QQ plot of the test (Figure S8). The QQ plots showed that the false positive rate of the MFRG for detection of interaction is controlled.

A total of 242 pairs of genes which were derived from 113 genes showed significant evidence of epistasis influencing five traits with P-values < $2.92\times10^{-10}$ which were calculated using the MFRG model (Table S11). Of them formed a largest connected subnetwork (Figure 10) in which a subnetwork connecting genes CSMD1, ST20 and SHPK were also observed in Figure 9. We listed top 25 pairs of significantly interacted genes with five traits in Table 7. We observed the same pattern as that we observed for two traits: HDL and LDL. Again, we observed that pairs of SNPs between two genes jointly have significant interaction effects, but individually each pair of SNPs might make mild contributions to the interaction effects as shown in Table S12. There were a total of 6766 pairs of SNPs between genes CSMD1 and FOXO1. Table S12 listed 73 pairs of SNPs with P-values < 0.049. Majority of the 73 pairs of SNPs showed no strong evidence of interaction. However, they collectively demonstrated significant interaction influencing five traits.

**Fig. 10. Networks of 242 pairs of genes showing significant evidence of interactions as identified by MFRG.**

Table 7. P-values of top 25 pairs of significantly interacted genes with five traits.

| Gene 1 | Gene 2 | P-values | | | | | | | |
|---|---|---|---|---|---|---|---|---|---|
| | | Five Traits | | | HDL | LDL | SBP | DBP | TOTCHOL |
| | | MFRG | Pair-wise (min) | PCA | FRG | FRG | FRG | FRG | FRG |
| *CSMD1* | *ICE2* | 5.03E-35 | 9.54E-04 | 5.94E-08 | 1.17E-04 | 1.47E-04 | 8.39E-03 | 1.32E-03 | 4.81E-04 |
| *PDZK1IP1* | *CSMD1* | 1.02E-32 | 2.42E-14 | 2.97E-10 | 3.34E-03 | 5.62E-02 | 5.64E-03 | 3.26E-03 | 6.37E-04 |
| *CSMD1* | *FOXO1* | 3.30E-30 | 1.50E-17 | 9.78E-08 | 8.26E-07 | 1.81E-06 | 3.11E-06 | 4.08E-07 | 5.08E-07 |
| *MEGF6* | *CSMD1* | 7.53E-29 | 1.51E-09 | 4.78E-54 | 1.44E-06 | 3.02E-04 | 1.00E-05 | 1.95E-06 | 6.83E-07 |
| *ICE2* | *ST20* | 1.36E-28 | 2.06E-03 | 3.40E-03 | 8.49E-05 | 2.67E-04 | 7.10E-05 | 2.40E-05 | 4.50E-05 |
| *ST20* | *SHPK* | 1.59E-28 | 7.94E-05 | 3.31E-02 | 2.12E-10 | 9.01E-09 | 2.23E-04 | 6.41E-05 | 4.93E-08 |
| *PDZK1IP1* | *ST20* | 2.14E-28 | 3.94E-14 | 1.57E-02 | 7.36E-03 | 4.73E-02 | 2.32E-03 | 6.29E-03 | 1.83E-03 |
| *CSMD1* | *SHPK* | 4.11E-28 | 5.74E-05 | 5.03E-05 | 2.31E-04 | 1.04E-03 | 3.87E-03 | 3.00E-02 | 1.11E-03 |
| *CSMD1* | *CCNDBP1* | 1.10E-27 | 1.64E-04 | 8.11E-03 | 2.98E-04 | 6.57E-03 | 7.42E-05 | 6.36E-04 | 1.34E-03 |
| *CSMD1* | *C14orf37* | 1.65E-27 | 1.39E-21 | 4.37E-11 | 1.30E-06 | 6.29E-05 | 1.70E-06 | 4.27E-05 | 1.45E-06 |
| *PAIP2B* | *CSMD1* | 3.70E-27 | 5.49E-07 | 9.74E-13 | 2.11E-06 | 5.40E-06 | 2.63E-09 | 7.39E-08 | 2.34E-05 |
| *C5orf64* | *SPDYC* | 5.51E-27 | 1.11E-07 | 3.93E-02 | 1.08E-04 | 2.81E-07 | 7.12E-05 | 5.93E-06 | 5.43E-04 |
| *C5orf64* | *SHPK* | 1.19E-26 | 8.25E-05 | 2.58E-02 | 5.55E-09 | 1.31E-08 | 2.55E-04 | 4.25E-06 | 7.48E-08 |
| *MEGF6* | *IRF2BPL* | 3.75E-26 | 2.45E-21 | 2.85E-15 | 7.78E-01 | 1.86E-01 | 3.29E-03 | 5.52E-02 | 1.72E-01 |
| *ROR2* | *IRF2BPL* | 1.17E-25 | 1.89E-21 | 3.61E-15 | 7.78E-02 | 3.88E-02 | 2.04E-01 | 2.33E-01 | 2.11E-02 |
| *CSMD1* | *BRAP* | 6.22E-25 | 4.92E-05 | 4.78E-07 | 1.89E-04 | 3.26E-03 | 5.32E-02 | 3.16E-02 | 3.60E-03 |
| *CCNDBP1* | *ST20* | 1.02E-24 | 2.04E-03 | 5.22E-02 | 8.54E-05 | 2.60E-06 | 1.90E-07 | 1.79E-06 | 5.40E-05 |
| *CPSF3L* | *PDZK1IP1* | 1.34E-24 | 1.08E-11 | 1.21E-12 | 2.97E-02 | 9.20E-02 | 4.67E-03 | 2.54E-02 | 8.88E-03 |
| *CPSF3L* | *CSMD1* | 2.89E-24 | 3.08E-09 | 3.64E-07 | 3.36E-03 | 3.54E-02 | 3.77E-01 | 3.20E-02 | 2.67E-03 |
| *CSMD1* | *PPM1A* | 2.89E-24 | 5.45E-06 | 1.07E-03 | 1.42E-05 | 5.39E-06 | 2.29E-02 | 3.74E-06 | 2.73E-04 |
| *C5orf64* | *IRF2BPL* | 3.34E-24 | 1.79E-21 | 1.20E-02 | 6.16E-02 | 6.71E-03 | 6.80E-03 | 7.15E-03 | 2.34E-03 |
| *CSMD1* | *CDC14B* | 3.39E-24 | 4.74E-03 | 3.52E-02 | 1.52E-03 | 1.39E-05 | 1.56E-02 | 1.46E-02 | 1.49E-03 |
| *IGHM* | *ST20* | 5.23E-24 | 5.72E-05 | 1.85E-02 | 2.05E-05 | 1.26E-07 | 1.30E-03 | 4.35E-05 | 7.44E-05 |
| *STX12* | *IGH* | 5.32E-24 | 1.12E-17 | 9.25E-07 | 2.80E-02 | 2.04E-01 | 3.18E-01 | 3.67E-02 | 1.62E-02 |
| *GSK3B* | *CSMD1* | 5.33E-24 | 8.66E-08 | 1.53E-03 | 1.26E-03 | 2.98E-06 | 6.78E-02 | 9.19E-03 | 2.50E-04 |

Among 242 significantly interacted genes identified by joint analysis with five traits, we observed only one pair of genes: ST20 and SHPK showed epistasis influencing LDL at the genome-wide significance level by univariate analysis of epistasis with the LDL individually. However, if we release the significance level to $P < 5.0 \times 10^{-8}$, we observed 7 pairs of genes showing pleiotropic epistasis effects by univariate of epistasis analysis individually (Figure 11). This demonstrated that although by each individual trait analysis, they only showed mild evidence of epistasis, by simultaneous epistasis analysis of multiple correlated traits the genes showed strong evidence of epistasis influencing multiple traits. The results imply that the genetic analysis of multiple traits can reveal the complicated genetic structures of the complex traits which may be missed by univariate genetic analysis.

**Fig. 11. Pleiotropic epistasis of HDL, LDL, DBP, SBP and TOTCHOL.**

## Discussion

Most genetic analyses of phenotypes have focused on analyzing single traits or, analyzing each phenotype independently. However, multiple phenotypes are highly correlated. Genetic variants can be associated with more than one trait. Genetic pleiotropic effects likely play a crucial role in the molecular basis of correlated phenotypes. To address these central themes and critical barriers in interaction analysis of multiple phenotypes, we shift the paradigm of interaction analysis from individual interaction analysis to pleiotropic interaction analysis and uncover the global organization of biological systems. We used MFRG to develop a novel statistical framework for joint interaction analysis of multiple correlated phenotypes. By large simulations and real data analysis we demonstrate the merits and limitations of the proposed new paradigm of joint interaction analysis of multiple phenotypes.

The new approach fully uses all phenotype correlation information to jointly analyze interaction of multiple phenotypes. By large simulations and real data analysis, we showed that the proposed MFRG for joint interaction analysis of correlated multiple phenotypes substantially increased the power to detect interaction while keeping the type 1 error rates of the test statistics under controls. In real data analysis, we observed that although pairs of genes showed no strong evidence of interactions influencing individual trait, they indeed demonstrated significant interactions if interactions were simultaneously analyzed for correlated multiple traits.

Due to lack of power of the widely used statistics for testing interaction between loci and its computational intensity, exploration of genome-wide gene-gene interaction has been limited. Few significant interaction results have been observed. Many geneticists question the universe presence of significant gene-gene interaction. Our analysis showed that although number of significantly interacted genes for single phenotype was small, the number of significantly interacted genes for multiple phenotypes substantially increased. Our results suggested that joint interaction analysis of multiple phenotypes should be advocated in the future genetic studies of complex traits.

The interaction analysis for multiple phenotypes has been limited to common variants in carefully controlled experimental crosses and has mainly focused on the pair-wise interaction analysis. Although pair-wise interaction analysis is suitable for common variants, but is difficult to use to test interaction between rare and rare variants, and rare and common variants. There is an increasing need to develop statistics that can be used to test interaction among the entire allelic spectrum of variants for joint interaction analysis of multiple phenotypes. The MFRG utilizes the merits of taking genotype as functions and decomposes position varying genotype function into orthogonal eigenfunctions of genomic position. Only a few eigenfunctions that capture major information on genetic variation across the gene, are used to model the genetic variation. This substantially reduces the dimension in genetic variation of the data. The MFRG can efficiently test the interaction between rare and rare, rare and common, and common and common variants.

In both real data analysis of two phenotypes and five phenotypes, the interacted genes formed interaction networks. We also observed the hub genes in the interaction networks. These hub genes usually play an important biological role in causing phenotype variation.

An essential issue for interaction analysis of a large number of phenotypes is how to reduce dimension while fully exploiting complementary information in multiple phenotypes. The standard multivariate regression models for joint interaction analysis of multiple phenotypes do not explore the correlation structures of multiple phenotypes and reduce the dimensions of the phenotypes, and hence have limited power to detect pleotropic interaction effects due to large degrees of freedom. Data reduction techniques such as principal component analysis should be explored in the future interaction analysis of multiple phenotypes.

The results in this paper are preliminary. The current marginal approaches for interaction analysis cannot distinguish between direct and indirect interactions, which will decrease our power to unravel mechanisms underlying complex traits. To overcome these limitations, causal inference tools should be explored for the joint interaction analysis of multiple phenotypes. The purpose of this paper is to stimulate further discussions regarding great challenges we are facing in the interaction analysis of high dimensional phenotypic and genomic data produced by modern sensors and next-generation sequencing.

## Acknowledgments

Mr. Xiong was supported by Grant 1R01AR057120–01 and 1R01HL106034-01, from the National Institutes of Health and NHLBI. Ms.

The authors wish to acknowledge the support of the National Heart, Lung, and Blood Institute (NHLBI) and the contributions of the research institutions, study investigators, field staff and study participants in creating this resource for biomedical research. Funding for GO ESP was provided by NHLBI grants RC2 HL-103010 (HeartGO), RC2 HL-102923 (LungGO) and RC2 HL-102924 (WHISP). The exome sequencing was performed through NHLBI grants RC2 HL-102925 (BroadGO) and RC2 HL-102926 (SeattleGO).

## Supporting Information

**S1 Fig. Power curves under Threshold with two genes including rare variants only.**

**S2 Fig. Power curves under Recessive OR Recessive with two genes including rare variants only.**

**S3 Fig. Power curves under Threshold with two genes including common variants only.**

**S4 Fig. Power curves under Recessive OR Recessive with two genes including common variants only.**

**S5 Fig. Power curves of MFRG with different sample size under Dominant AND Dominant.**

**S6 Fig. Power curves of MFRG with different sample size under Recessive OR Recessive.**

**S7 Fig. Power curves of MFRG with different sample size under Threshold.**

**S8 Fig. QQ plot of the five-trait FRG test for ESP dataset..**

**S1 Table. Average type 1 error rates of the statistic for testing interaction between two genes with no marginal effect consisting only rare variants with 2 traits over 10 pairs of genes.**

**S2 Table. Average type 1 error rates of the statistic for testing interaction between two genes with no marginal effect consisting only rare variants with 10 traits over 10 pairs of genes.**

**S3 Table. Average type 1 error rates of the statistic for testing interaction between two genes with marginal effect consisting only rare variants with 2 traits over 10 pairs of genes.**

**S4 Table. Average type 1 error rates of the statistic for testing interaction between two genes with marginal effect at one gene consisting only rare variants with 10 traits over 10 pairs of genes.**

**S5 Table. Average type 1 error rates of the statistic for testing interaction between two genes with marginal effects at two genes consisting only rare variants with 2 traits over 10 pairs of genes.**

**S6 Table. Average type 1 error rates of the statistic for testing interaction between two genes with marginal effects at two genes consisting only rare variants with 10 traits over 10 pairs of genes.**

**S7 Table. Average type 1 error rates of the statistic for testing interaction between two genes with marginal effects at two genes consisting only common variants with 2 traits over 10 pairs of genes.**

**S8 Table. Average type 1 error rates of the statistic for testing interaction between two genes with marginal effects at two genes consisting only common variants with 10 traits over 10 pairs of genes.**

**S9 Table. The interaction models: 0 and r stand for a quantitative trait mean given the genotypes.**

**S10 Table. P-values of significantly interacted genes with HDL and LDL.**

**S11 Table. P-values of significantly interacted genes with HDL, LDL, SBP, DBP and TOTCHOL.**

**S12 Table. P-values of 73 pairs of SNPs between genes CSMD1 and FOXO1 for testing interaction affecting five traits.**

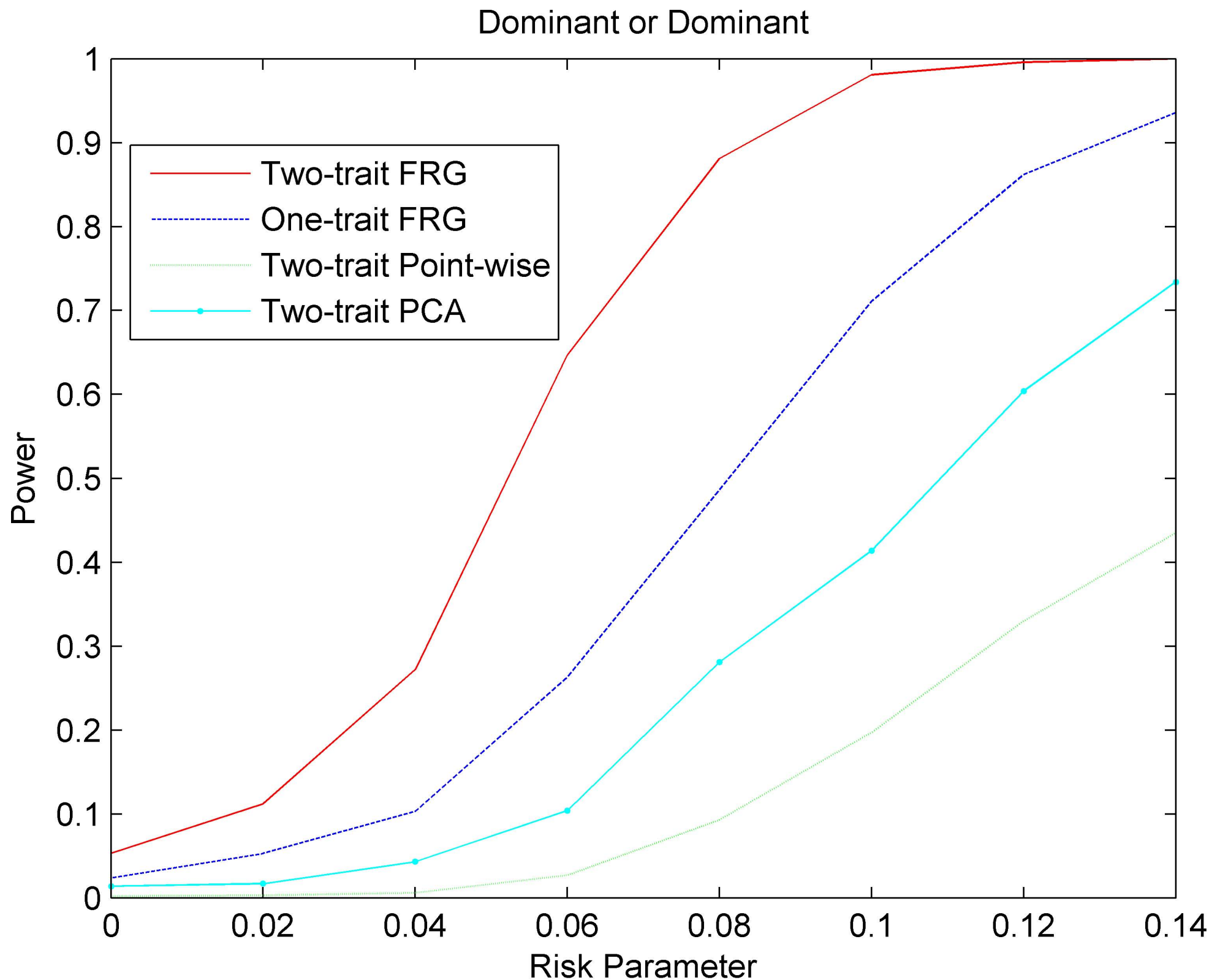
Dominant or Dominant
Two-trait FRG
One-trait FRG
Two-trait Point-wise
Two-trait PCA
Power
Risk Parameter
0
0.1
0.2
0.3
0.4
0.5
0.6
0.7
0.8
0.9
1
0
0.02
0.04
0.06
0.08
0.1
0.12
0.14

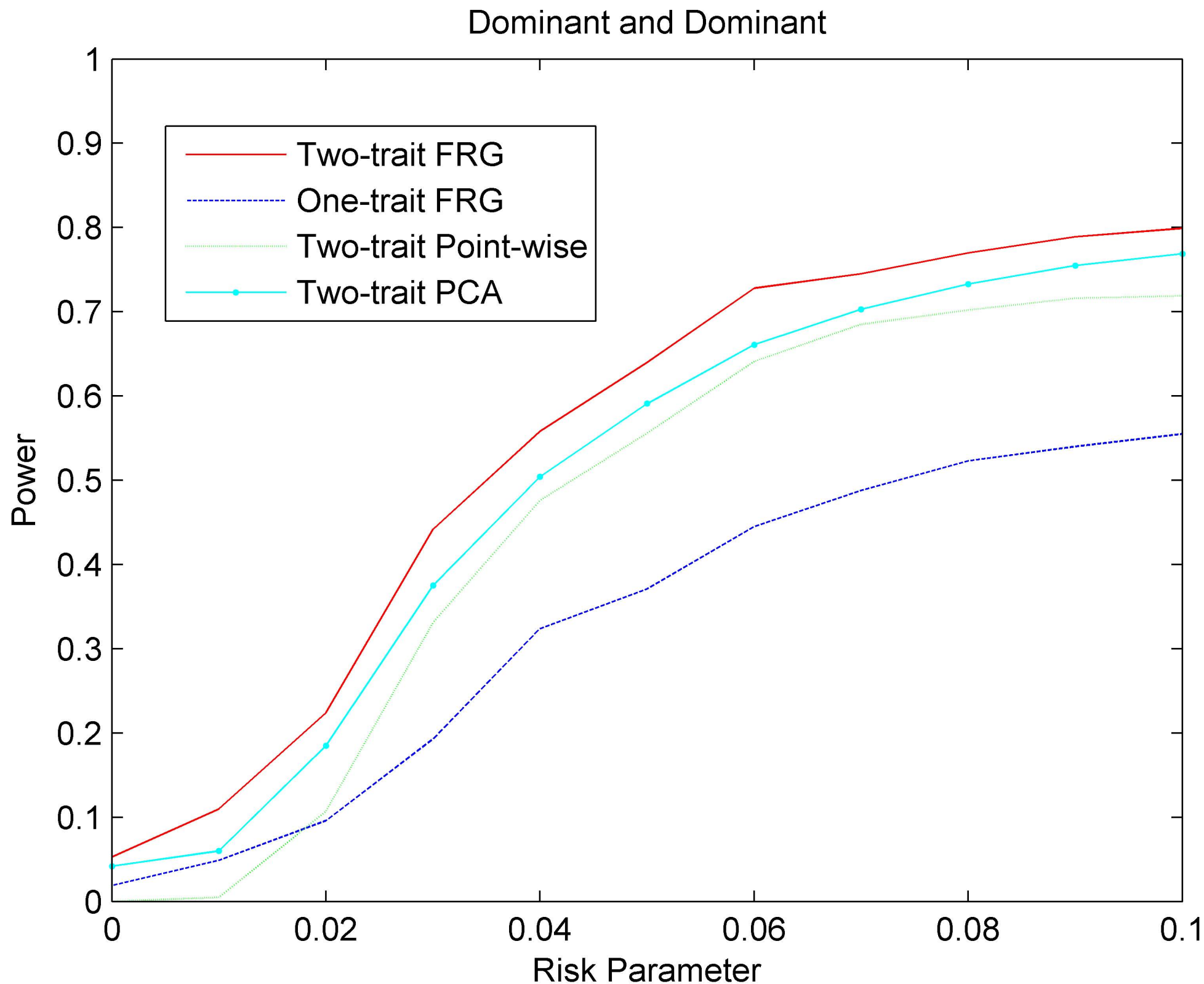

Dominant and Dominant
Two-trait FRG
One-trait FRG
Two-trait Point-wise
Two-trait PCA
Power
Risk Parameter
0
0.1
0.2
0.3
0.4
0.5
0.6
0.7
0.8
0.9
1
0
0.02
0.04
0.06
0.08
0.1

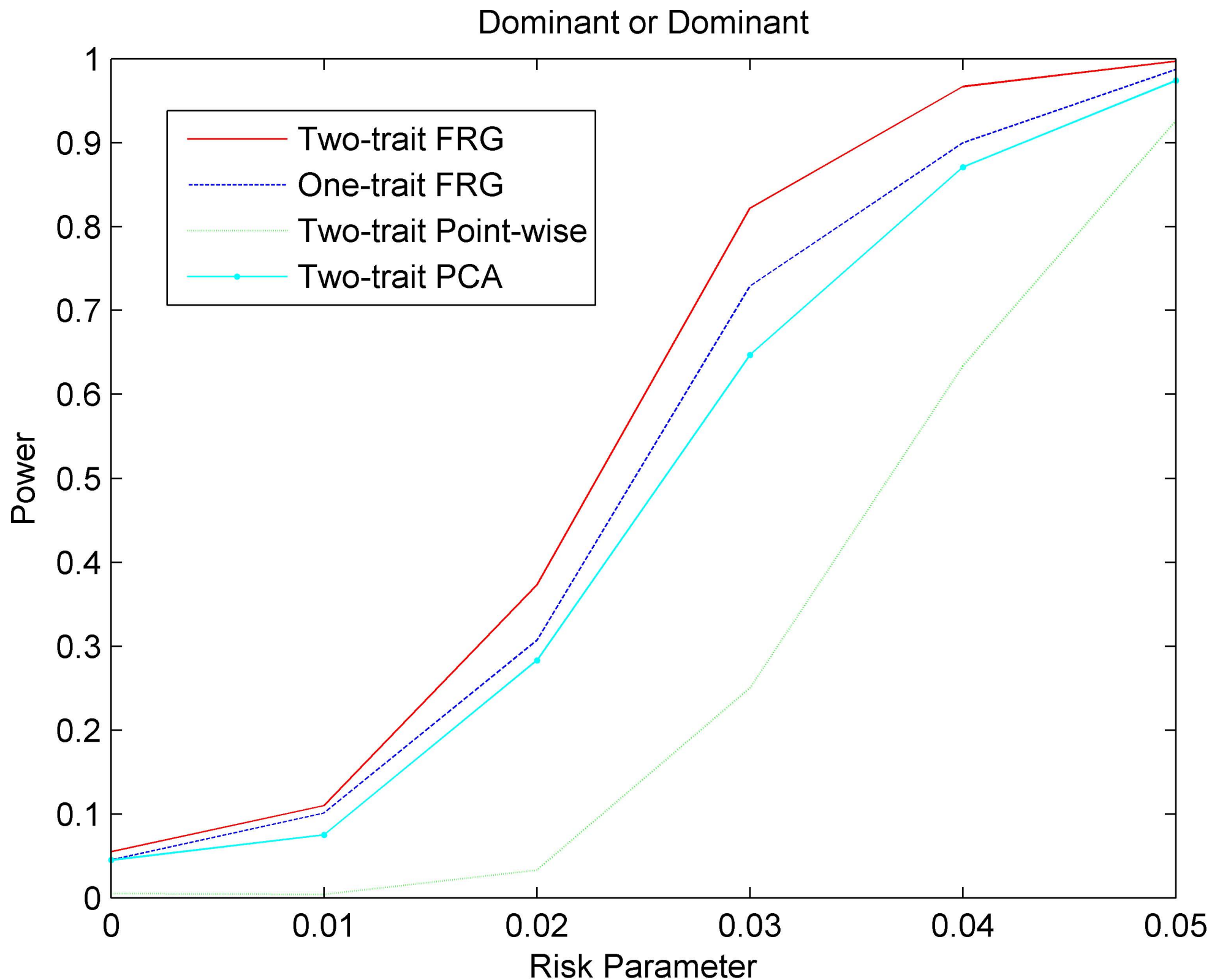
Dominant or Dominant
Two-trait FRG
One-trait FRG
Two-trait Point-wise
Two-trait PCA
Power
Risk Parameter
0
0.1
0.2
0.3
0.4
0.5
0.6
0.7
0.8
0.9
1
0
0.01
0.02
0.03
0.04
0.05

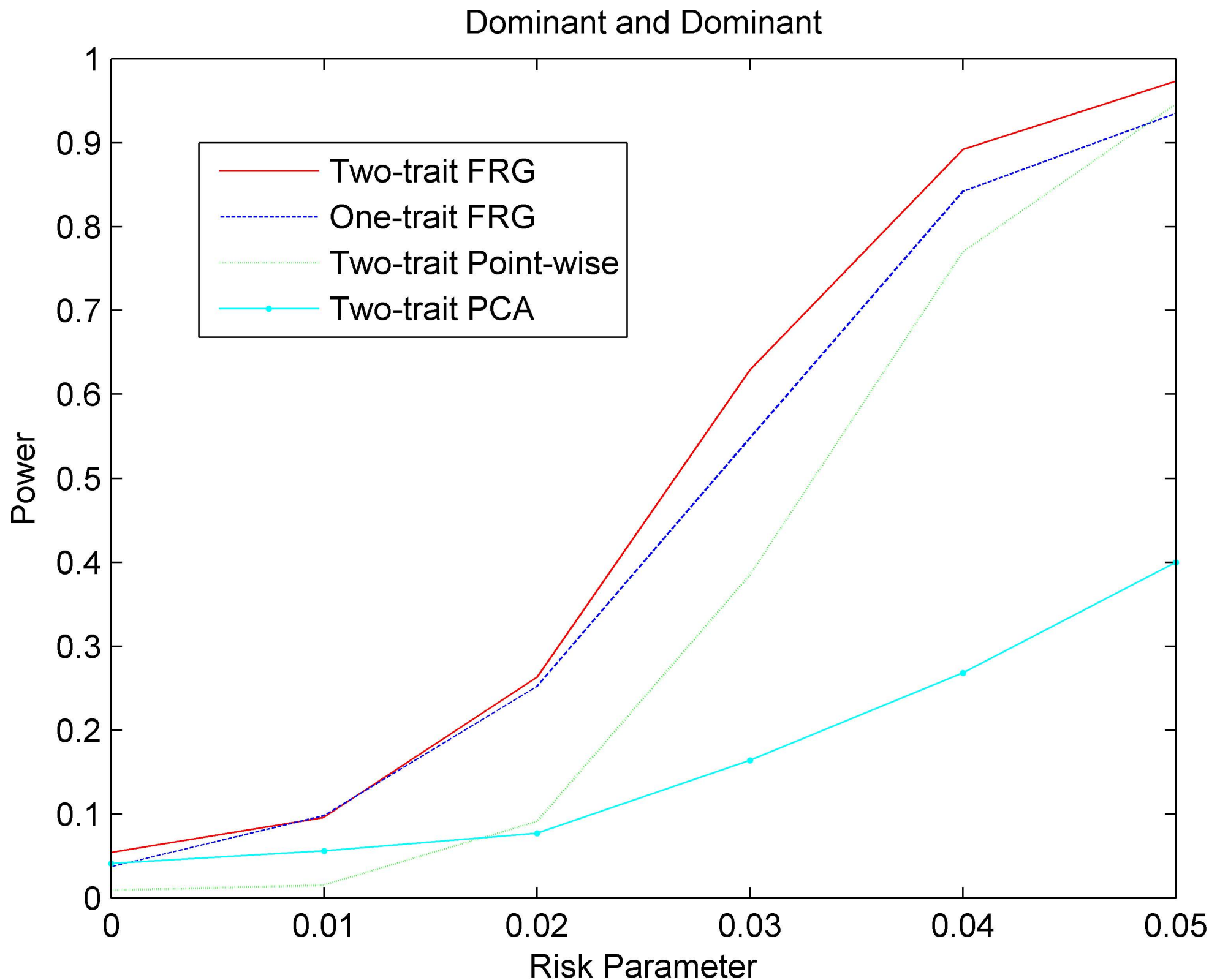
Dominant and Dominant
Two-trait FRG
One-trait FRG
Two-trait Point-wise
Two-trait PCA
Power
Risk Parameter
0
0.1
0.2
0.3
0.4
0.5
0.6
0.7
0.8
0.9
1
0
0.01
0.02
0.03
0.04
0.05

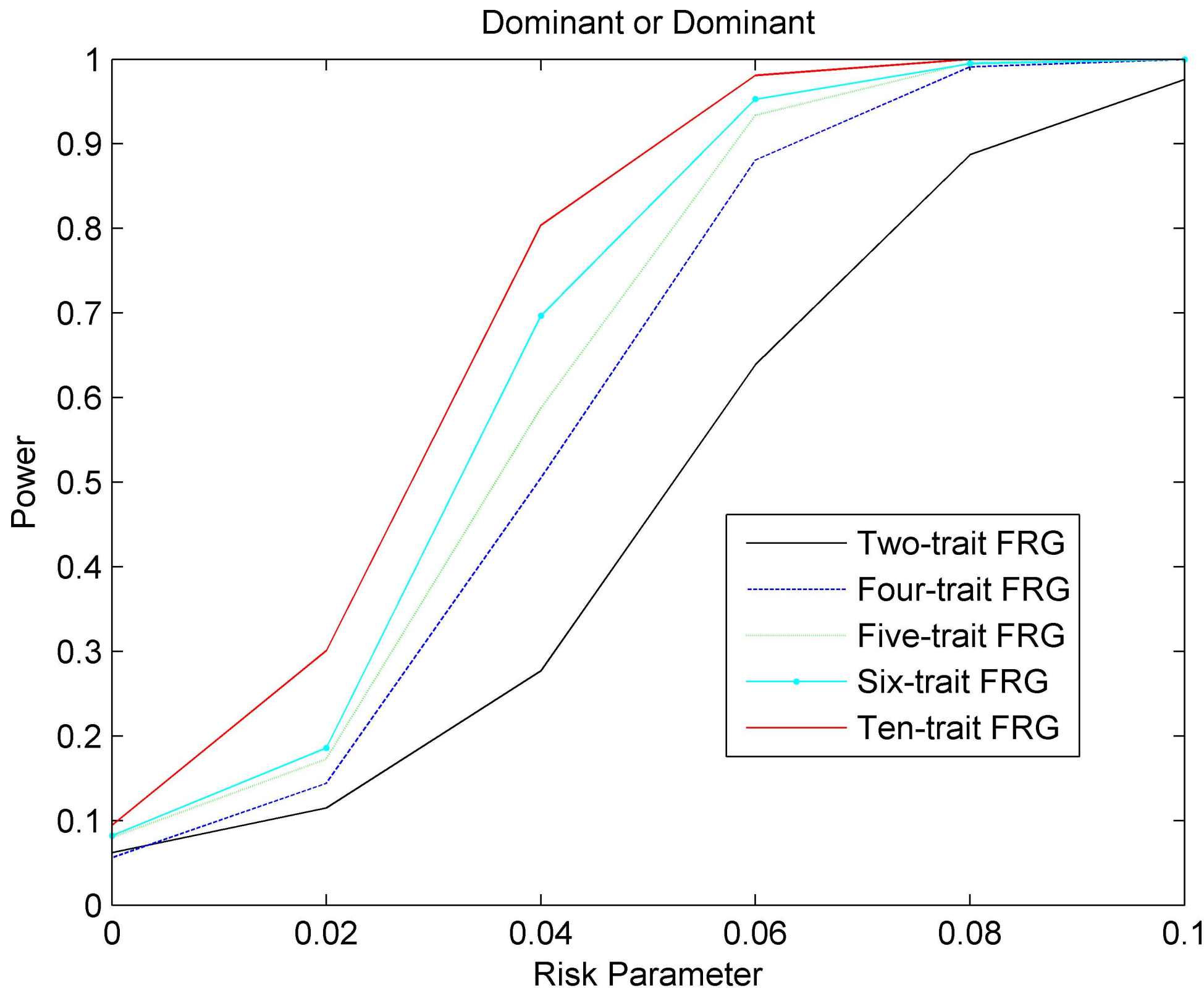
Dominant or Dominant
Power
Risk Parameter
0
0.1
0.2
0.3
0.4
0.5
0.6
0.7
0.8
0.9
1
0
0.02
0.04
0.06
0.08
0.1
Two-trait FRG
Four-trait FRG
Five-trait FRG
Six-trait FRG
Ten-trait FRG

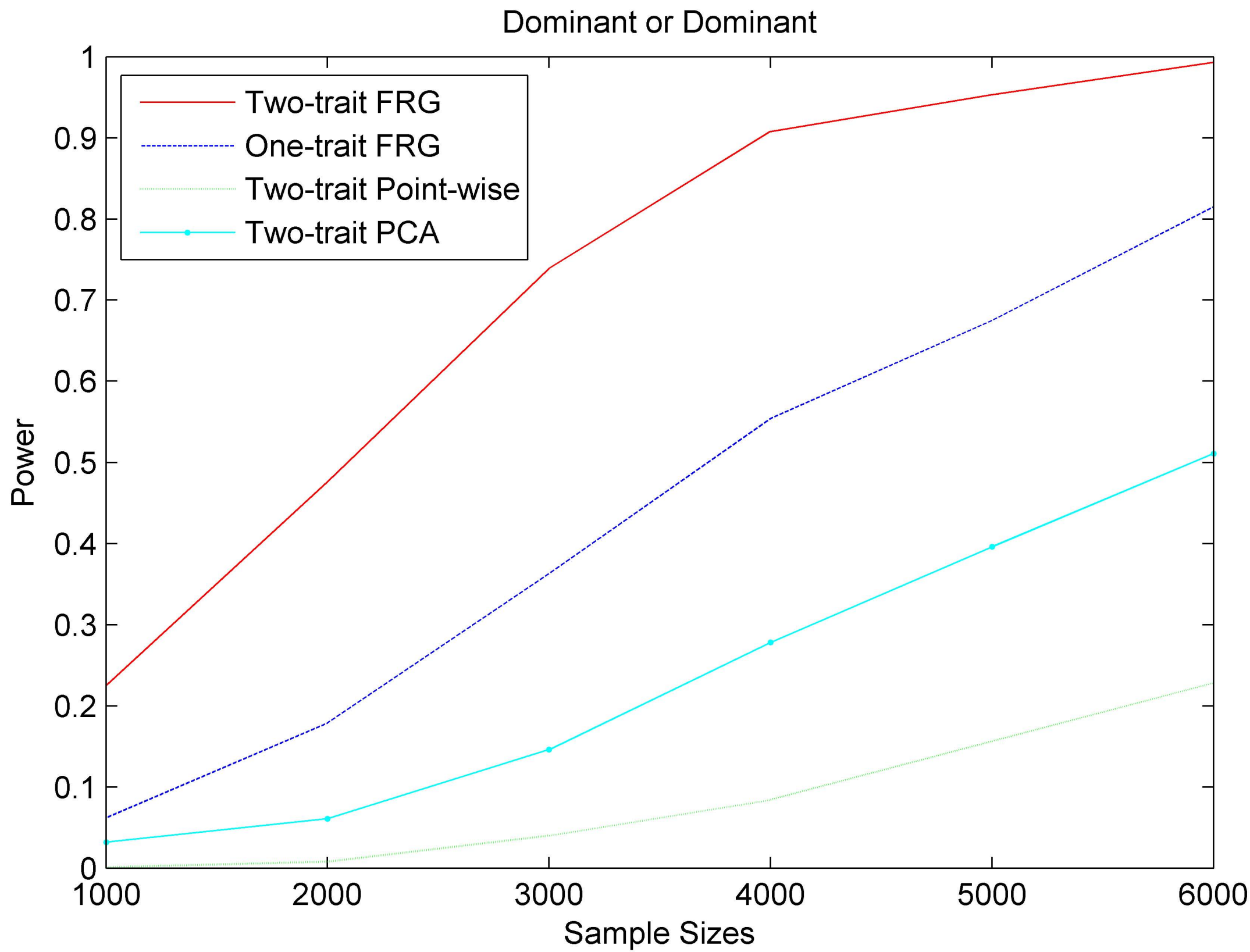
Dominant or Dominant
Two-trait FRG
One-trait FRG
Two-trait Point-wise
Two-trait PCA
Power
Sample Sizes
0
0.1
0.2
0.3
0.4
0.5
0.6
0.7
0.8
0.9
1
1000
2000
3000
4000
5000
6000

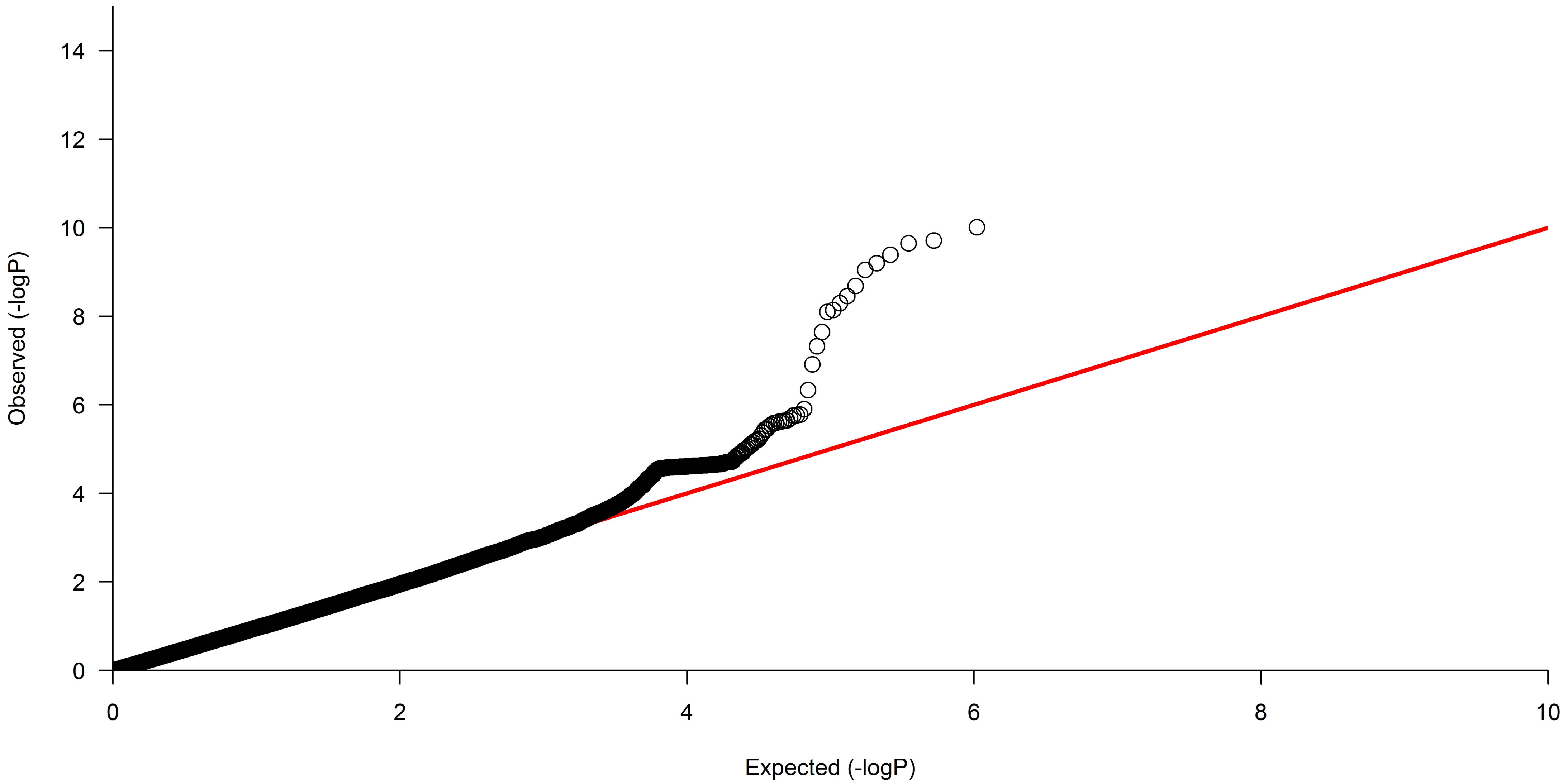

Observed (-logP)
0
2
4
6
8
10
12
14
0
2
4
6
8
10
Expected (-logP)

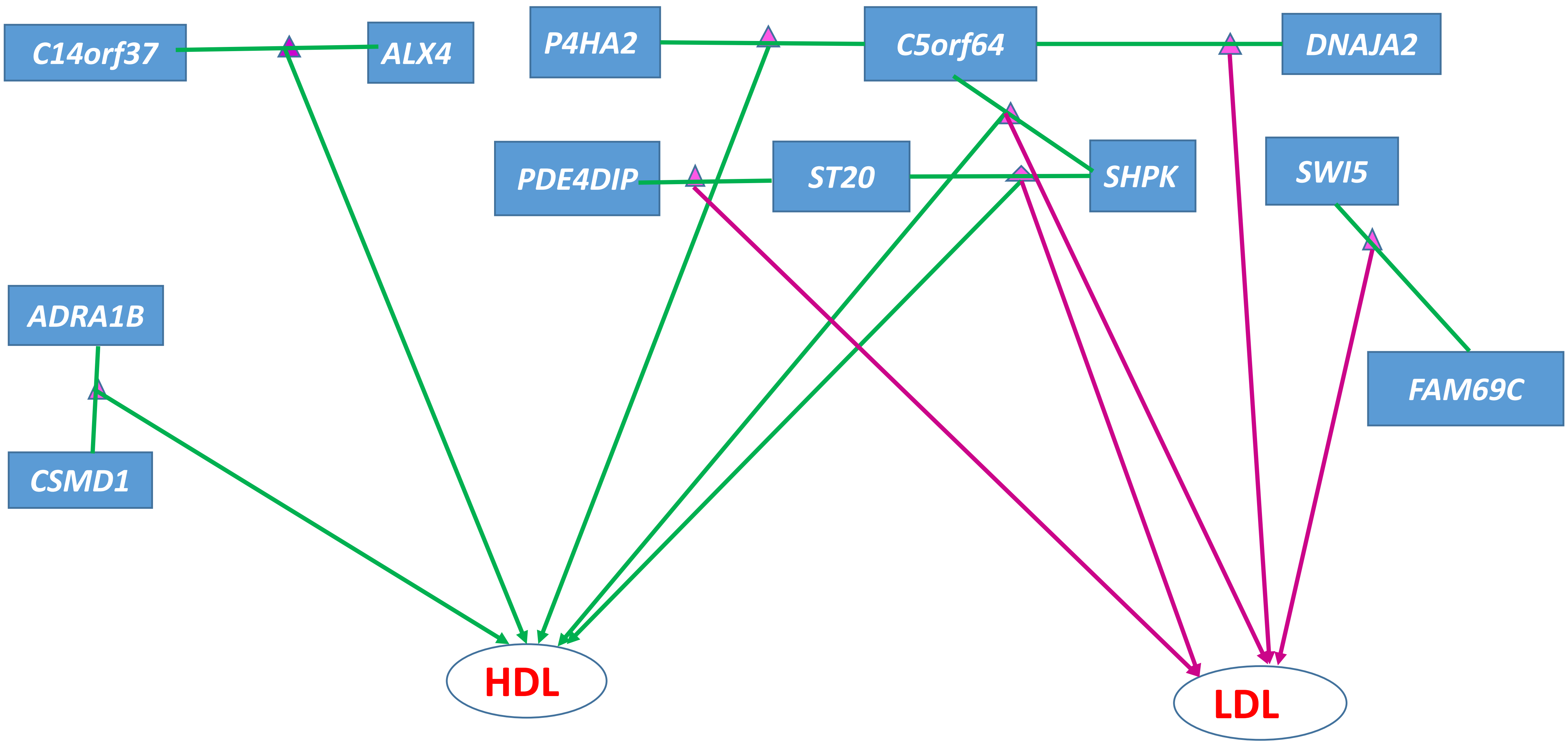

C14orf37
ALX4
P4HA2
C5orf64
DNAJA2
PDE4DIP
ST20
SHPK
SWI5
ADRA1B
FAM69C
CSMD1
HDL
LDL

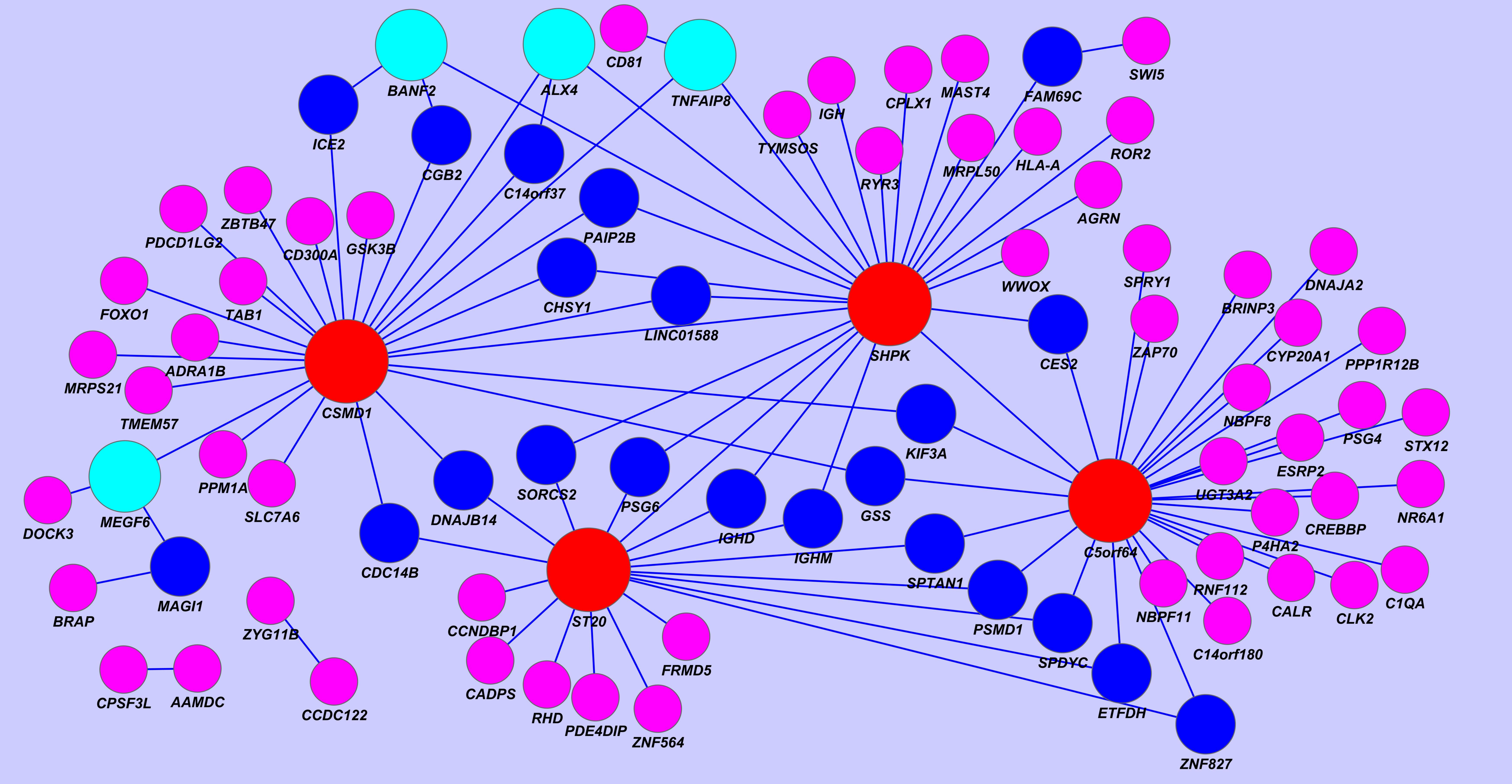
BANF2
ALX4
CD81
TNFAIP8
ICE2
CGB2
C14orf37
TYMSOS
IGH
CPLX1
MAST4
FAM69C
SWI5
RYR3
MRPL50
HLA-A
ROR2
AGRN
ZBTB47
PDCD1LG2
CD300A
GSK3B
PAIP2B
FOXO1
TAB1
CHSY1
LINC01588
SHPK
WWOX
SPRY1
DNAJA2
BRINP3
ZAP70
CES2
CYP20A1
PPP1R12B
ADRA1B
MRPS21
TMEM57
CSMD1
NBPF8
KIF3A
PSG4
STX12
ESRP2
UGT3A2
MEGF6
PPM1A
SLC7A6
DOCK3
DNAJB14
SORCS2
PSG6
IGHD
IGHM
GSS
CREBBP
NR6A1
P4HA2
C5orf64
CDC14B
MAGI1
BRAP
ZYG11B
SPTAN1
RNF112
CALR
CLK2
C1QA
NBPF11
PSMD1
C14orf180
CCNDBP1
ST20
FRMD5
CADPS
RHD
PDE4DIP
ZNF564
SPDYC
ETFDH
ZNF827
CPSF3L
AAMDC
CCDC122

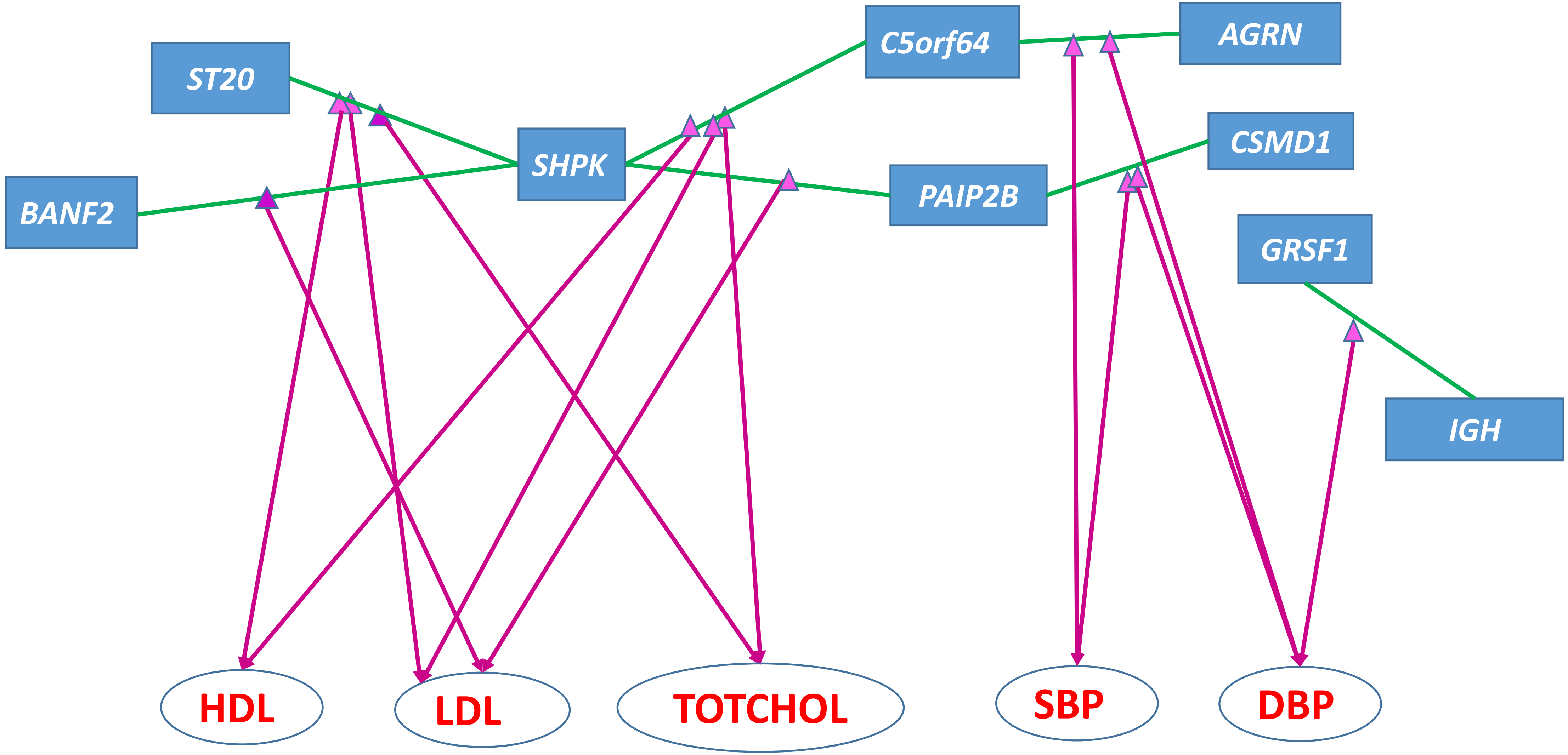

ST20
BANF2
SHPK
C5orf64
PAIP2B
AGRN
CSMD1
GRSF1
IGH
HDL
LDL
TOTCHOL
SBP
DBP